\documentclass[twocolumn,secnumarabic,amssymb, nobibnotes, aps, prd, preprintnumbers]{revtex4}
\usepackage{graphicx}
\usepackage{color}
\usepackage{fancyhdr}
\def\lsim{\mathrel{\rlap{\lower4pt\hbox{\hskip1pt$\sim$}}
    \raise1pt\hbox{$<$}}}         %less than or approx. symbol
\def\gsim{\mathrel{\rlap{\lower4pt\hbox{\hskip1pt$\sim$}}
    \raise1pt\hbox{$>$}}}         %greater than or approx. symbol
\begin{document}
\preprint{IPMU09-0119}

%%%%%%%%%%%%%%%%%%%%%%%%%%%%%%%%%%%%%%%%%%%%
%% FRONTMATTER
%%%%%%%%%%%%%%%%%%%%%%%%%%%%%%%%%%%%%%%%%%%%

%\begin{document}

\title{Dark Matter and Collider Physics in Split-UED \footnote{For
proceedings of SUSY09, the 17th international conference on Supersymmetry and the Unification
of Fundamental Interactions at Northeastern University, Boston, U.S.A. in 5-10 June, 2009.}}

%\classification{11.10.Kk, 95.35.+d, 98.70.Rz}
%\keywords      {dark matter, cosmic ray, extra dimension, Kaluza-Klein parity, LHC}

\author{Seong Chan Park, Jing Shu}
\affiliation{
Institute for the Physics and Mathematics of the Universe, University of Tokyo, Kashiwa, Chiba 277-8568, JAPAN}

\begin{abstract}
Kaluza-Klein dark matter is an attractive weakly interacting massive particle in universal extra dimension model. In the recent extension "split-UED", annihilation of Kaluza-Klein dark matter with a mass range $600-1000$ GeV provides excellent fits to the recently observed excesses in cosmic electron and positron fluxes of Pamela, ATIC and Fermi-LAT experiments. The cosmic gamma-ray flux in the same process can be significant around 300 GeV, thus can be observed or constrained by the forthcoming Fermi-LAT diffuse gamma-ray data. The collider signal at the LHC is the resonance in the dijets channels and the large missing energy in the missing energy plus jets. 
\end{abstract}

%the relic density of Kaluza-Klein dark matter with $600-1000$ GeV mass is in good agreement with the observed dark matter amount in our universe.   Interestingly the annihilation of Kaluza-Klein dark matter in the same mass range also provides excellent fits to the recently observed excesses in cosmic electron and positron fluxes of Pamela, ATIC and Fermi-LAT experiments. 

\maketitle

%%%%%%%%%%%%%%%%%%%%%%%%%%%%%%%%%%%%%%%%%%%%
%% MAINMATTER
%%%%%%%%%%%%%%%%%%%%%%%%%%%%%%%%%%%%%%%%%%%%

\section{Model: UED and split-UED}
Universal Extra Dimension (UED) \cite{UED} and its recent extension split-UED \cite{sued1} are flat extra dimension models based on a $S^1/Z_2$ orbifold. All standard model fields are universally propagating through the 5D bulk so that their Kaluza-Klein (KK) excited states are phenomenologically important if the compactification scale is low ($1/R\sim {\rm TeV}$). The model has an underlying symmetry, called the KK parity, that prevents the lightest KK odd state to decay. If such a state is neutral, then it provides a viable dark matter candidate. In the minimal setup, the gauge group is the same as the standard model one, $G={\rm SU(3)_c\times SU(2)_W \times U(1)_Y}$. Quarks and leptons are also alleviated to higher dimensional ones and their KK spectra are doubled. One chirality of their zero modes is projected out by the orbifold condition, leaving the other one as the standard model fermions. The most prominent feature of split-UED is the presence of (double) kink masses for fermions while keeping the KK parity. The five dimensional action of split-UED is the same as the minimal UED (MUED) except the 5D bulk mass terms:
\begin{eqnarray}
S_{\rm split-UED} = S_{\rm mUED} - \int d^4 x \int_{-L}^{L} \sqrt{g} \,\,m^{ij}_5(y)\bar{\Psi}_i \Psi_j
\end{eqnarray}
where $\Psi_i$ are the bulk Dirac spinors containing quarks and leptons in their zero modes and the kink mass, $m^{ij}_5(y) = \mu^{ij}_5 \theta(y)$, is introduced with a step function defined as $\theta(y>0)=1$ and $\theta(y<0)=-1$. Here $\pm$ signs depends on the the chirality of the zero mode.

Once 5D bulk mass is introduced, KK fermions get additional masses and become heavier \begin{eqnarray}
m_n^2 = m_0^2  + k_n^2 + \mu_5^2
\end{eqnarray}
where the first term, second term and third term are coming from the ordinary SM Yukawa interaction, the momentum of the extra dimension and the 5D bulk mass, respectively. $k_n$ is determined by $k_n L = n\pi$ for KK even modes and the n-th solution of $\mu_5 = \pm k_n \cot k_n L$ for KK odd modes. 

Having the generic idea of split-UED, we can control the KK spectra by turning on some bulk mass parameters. In order to suppress the hadronic annihilation cross section and avoid the unnecessary flavor problems, we first choose our 5D bulk masses for quarks to be universal and larger than the typical KK scale, which is $\mu^{ij}_q = \mu_5 \,\delta_{ij}$. For the charged leptons, in order to control their annihilation cross section ratio among different flavors which are dominantly coming from the right-handed components, we choose separate 5D bulk masses $\mu_{e_R}$, $\mu_{\mu_R}$ and $\mu_{\tau_R}$ to achieve that. We left the bulk masses of the left-handed leptons to be small to evade the experimental constrains from the lepton flavor violation and four fermion operators. 

%For the the left-handed leptons, we might assume that their 5D bulk mass are generally small so that their couplings are still almost KK conserving. Therefore their contributions to the lepton flavor violation and four fermion operators are negligible.

\section{Fitting Pamela, ATIC and Fermi-LAT}

The KK dark matter ($B_1$), mainly annihilates into fermion pairs among which charged lepton pair is dominant. Stable particles, like proton (p), electron ($e^-$) and photon ($\gamma$), and their antiparticles are generated from the casecade decay of hadrons and heavy charged leptons in the $B_1 B_1$ annihilation process and propagate to the Earth. In \cite{sued2, sued3} we calculate the electron, positron and gamma-ray flux from the annihilation of $B_1$ and compare them with the recent observations from PAMELA \cite{Pamela}, ATIC \cite{ATIC} and Fermi-LAT collaborations \cite{Fermi-LAT}.

\begin{figure}[b]
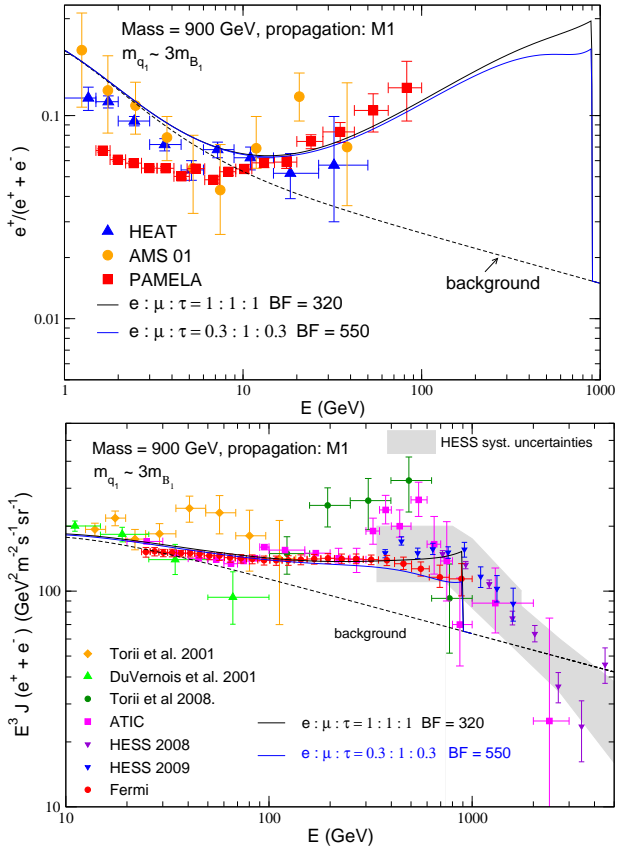

  \includegraphics[trim=0 0 0 0, clip, width=.45\textwidth]{pamela.eps}
  \includegraphics[trim=0 0 0 0, clip, width=.45\textwidth]{fermi.eps}
  \caption{{\it (Upper)} The positron fraction ($e^+/(e^+ + e^-)$) with $m_{B_1} = 900$ GeV, compared with PAMELA data. The mass of KK quark is taken to be about $3m_{B_1}$. The black curve is for $e:\mu:\tau=1:1:1$ case and the blue curve is for $0.3:1:0.3$ case. 
  {\it (Lower)} The total flux of positron and electron with $m_{B_1} = 900$ GeV. 
 \label{fig1}}
\end{figure}

In Fig ~\ref{fig1}, we fit the positron fraction ($e^+/ (e^-+e^+)$) in the Pamela energy range (upper) and electron plus positron flux ($e^+e^-$) in the Fermi-LAT range (lower) with $m_{B_1}=900$ GeV. For fitting the ATIC "peak and sharp drop-off pattern", a lower mass $m_{B_1}=620$ GeV is preferred as shown in \cite{sued2}. The anti-proton flux is not a problem in split-UED because the hadronic branching fraction is suppressed by heavy KK quark masses. In the figures we took $m_{q_1} = 3 m_{B_1}$ so that the hadronic branching fraction is less than $10 \%$. %We also want to point out is the smooth shape at the end point of electron plus positron flux can be achieved in split-UED by 5D bulk masses for leptons. %The KK quark mass spectra are taken to be $m_{q_1} \approx 3/R$ so that we are able to reach their KK quark production and decay signals at the LHC \cite{sued2}.

\begin{figure}[h]
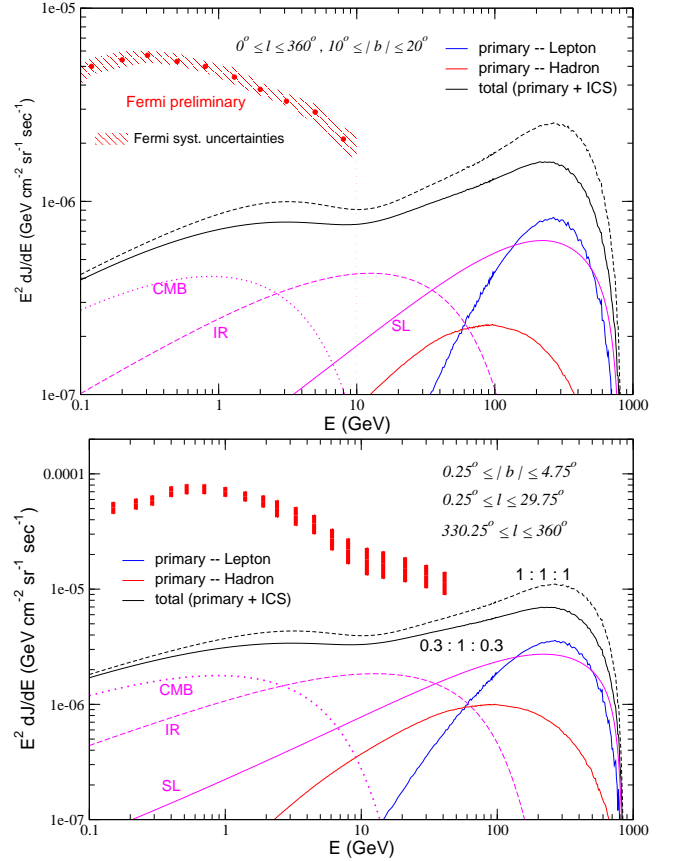
%[htbp]
  \includegraphics[width=.47\textwidth]{gamma1}\\
  \includegraphics[width=.47\textwidth]{gamma2}
  \caption{The gamma-ray signal from $B_1$ dark matter annihilation for the regions of Inner Galactic plane (\textit{Upper}) and intermediate galactic latitudes (\textit{Lower}). The black solid and dashed curves are the sum of primary and Inverse Compton Scattering (ICS) contributions when the final state charged lepton branching ratios are $(0.3:1:0.3)$ and $(1:1:1)$ respectively. The blue and red lines are the gamma-ray from $\tau^\pm$ and hadrons from $B_1$ dark matter annihilation, respectively, and the magenta lines are contributions of three components of ICS in the $(0.3:1:0.3)$ case.
\label{fig2}}
\end{figure}

In Fig.~\ref{fig2}, we show the gamma-ray signal for Inner Galactic plane region and the region of intermediate galactic latitudes. The black curve is the sum of primary and ICS contributions, the blue and red lines are the gamma-ray from $\tau^\pm$ and hadrons from $B_1$ dark matter annihilation, respectively, and the magenta lines are contributions of three components of ICS. In the region of low energy ($E \lsim10$ GeV), the dark matter signal is much smaller than the observed data, however, starting from few tens of GeV, the signal is about a factor of $2\sim3$ smaller. A characteristic of our model is that a bump at $E\approx 300$ GeV can be seen, due to the main contributions from primary $\tau^\pm$ and also the ICS from the star light in a subleading way, which can be checked soon if higher energy of gamma ray data is available. We also show the case of universa KK lepton in dashed line for reference. The behavior is similar to the non-universal case, but with more energetic gamma-ray due to more hard $e^\pm$ and more $\tau^\pm$. 

\section{Collider physics}
\begin{figure}[h]%[htbp]
  \includegraphics[width=.47\textwidth]{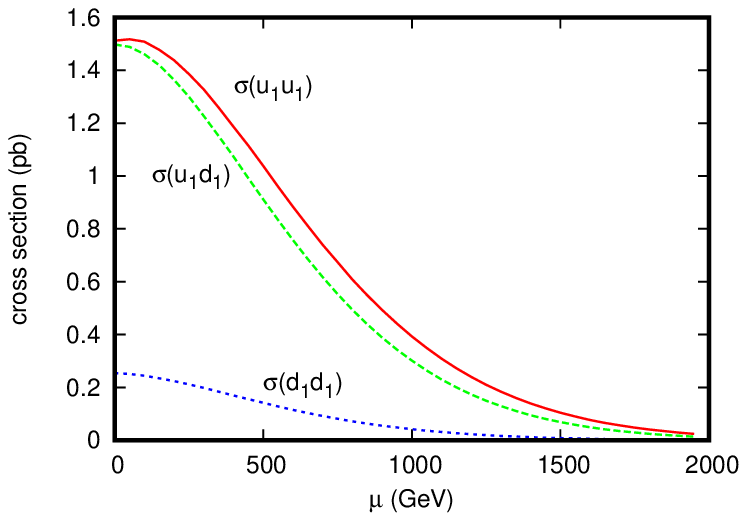}
  \includegraphics[width=.47\textwidth]{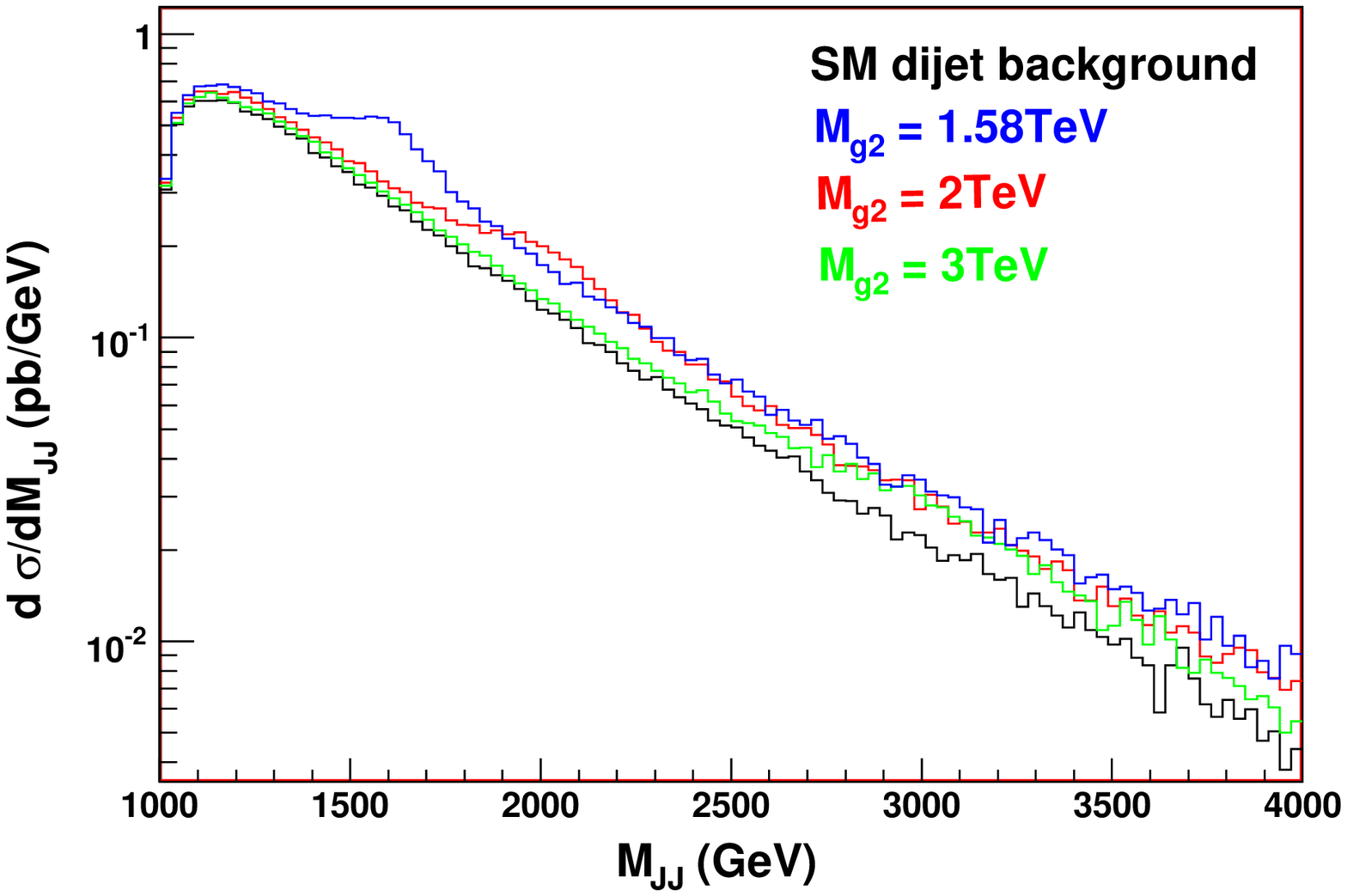}
  \caption{{\it (Upper)} The cross sections as functions of the bulk mass $\mu$ with 
 fixing $1/R=900$ GeV.
Here, $u_1$($d_1$) includes both $u_{L1}$ and $u_{R1}$ ($d_{L1}$ and $d_{R1}$) contributions. {\it (Lower)} The invariant mass distribution of the dijet signal. The resonance peaks appear at the
$M_{g^2}$ for 1.58 (for ATIC/PPB-BETS), 2 and 3 TeV.
\label{fig3}}
\end{figure}

At the LHC, the 1st KK colored particles (KK quarks and gluons) in split-UED are copiously produced and further decay into hard jets, soft jets/leptons and dark matter (missing energy), see Fig. \ref{fig3} (upper).  Because of the large mass splitting between the 1st KK quark $q_1$ and 1st KK gluon $g_1$, we can separate out the production of $g_1 g_1$, $g_1 q_1$ and $q_1 q_1$ by using the $M_{\textrm{eff}}$ cuts. We focus on the $q_1q_1$ production, which has two hard jets so that we can distinguish the signals from the SM background. The split-UED $q_1$ pair signal is simulated by scaling the SUSY signatures with the same mass spectrum. After the appropriate cuts, our signal is well above the SM background. We calculate the $M_{T2}$ distribution of the $q_1$ pair signals as shown in \cite{sued2}, and its end point reflects the information for combination of $q_1$ and $g_1$ masses. Although we are unable to determine the mass of individual particle in split-UED, its predictions of the signatures of $q_1q_1$ production at the LHC, e.g. $M_{\rm eff}$ and $M_{T2}$, can be examined whether they are consistent with the cosmic-ray signatures in the near future, and vice versa.   

Because of the KK number violation at the tree level, the KK even gauge bosons, especially the 2nd KK gluon ($g^2$), can be singly produced at a large rate. We expect to discover the resonance in the dijet channels against the large QCD background \cite{Lillie:2007ve} even in the early stages of LHC ($\mathcal{L}=$100 $ $pb$^{-1}$). To explore this possibility, in Fig.  \ref{fig3} (lower), we plot the invariant mass distribution of QCD dijets simulated by Madgraph \cite{Alwall:2007st} (with rough acceptance cuts $ |\eta| < 2.5$ and $p_T>500$ GeV to reduce the SM background). We have found that even at the early stage, one can discover the $g^2$ just below 4 TeV. For the LHC running time, a $g^2$ with its mass up to more than 6 TeV can be discovered \cite{sued1}.

%\section{Conclusion}

%We consider the pair annihilation of the lightest Kaluza-Klein photons in split-UED as a primary source of recently observed cosmic ray positron and gamma in PAMELA, ATIC and Fermi-LAT. Leptophilic property of dark matter suggested by the PAMELA antiproton data is naturally realized in split-UED. 

%As the mass of dark matter particle around $900$ GeV and its primary annihilation channel being lepton pairs with $e:\mu:\tau=0.3:1:0.3$ (or $1:1:1$) we successfully fit the all the cosmic ray data. A particularly interesting prediction of our model is that the excess of cosmic gamma-ray flux, if observed by the forthcoming data of Fermi-LAT diffuse gamma-ray, peaks at $E\approx300$ GeV range. If there is no excess in the high energy region, then Fermi-LAT will put an upper bound on the tau fraction in our model.

%Finally we point out another interesting prediction for the collider phenomenology. In the case of splitting right handed charged leptons (i.e. $0.3:1:0.3$ case) a large cross section of dilepton (in particular, $e_R e_R$ and $\tau_R \tau_R$) production is expected at the LHC through 2nd KK gauge boson exchanges. As these leptonic signals are rather clean we expect that the detection would be promising and we leave it for future study.

%%%%%%%%%%%%%%%%%%%%%%%%%%%%%%%%%%%%%%%%%%%%%%%%
%% BACKMATTER
%%%%%%%%%%%%%%%%%%%%%%%%%%%%%%%%%%%%%%%%%%%%%%%%

%\begin{theacknowledgments}
%{\bf Acknowledgments}
\begin{acknowledgements}
This paper is based on works \cite{sued1, sued2, sued3} with C.-R. Chen, M. Nojiri and M. Takeuchi. These works were supported by the World Premier International Research Center Initiative (WPI initiative) by MEXT, Japan. SCP and JS were supported by the Grant-in-Aid for scientific research (Young Scientists (B) 21740172) and (Young Scientists (B) 21740169) from JSPS, respectively.  
  %\end{theacknowledgments}
\end{acknowledgements}



\bibliographystyle{aipprocl} % if natbib is missing

\begin{thebibliography}{999}

\bibitem{UED}
  T.~Appelquist, H.~C.~Cheng and B.~A.~Dobrescu,
  %``Bounds on universal extra dimensions,''
  Phys.\ Rev.\  D {\bf 64}, 035002 (2001).
%  [arXiv:hep-ph/0012100].
  %%CITATION = PHRVA,D64,035002;%%

\bibitem{sued1}
  S.~C.~Park and J.~Shu,
%  ``Split-UED and Dark Matter,''
  Phys.\ Rev.\  D (R) {\bf 79}, 091702 (2009). % {\it Rapid Communication}
 % [arXiv:0901.0720 [hep-ph]].
  %%CITATION = PHRVA,D79,091702;%%
  
%  \bibitem{Antoniadis:1990ew}
 % I.~Antoniadis,
%  ``A Possible new dimension at a few TeV,''
 % Phys.\ Lett.\  B {\bf 246}, 377 (1990).
  %%CITATION = PHLTA,B246,377;%% 
  
  \bibitem{sued2}
   C.~R.~Chen, M.~M.~Nojiri, S.~C.~Park, J.~Shu and M.~Takeuchi,
%  ``Dark matter and collider phenomenology of split-UED,''
  JHEP {\bf 0909}, 078 (2009).
%  [arXiv:0903.1971 [hep-ph]].
  
  \bibitem{sued3}
   C.~R.~Chen, M.~M.~Nojiri, S.~C.~Park and J.~Shu,  
%  ``Kaluza-Klein Dark Matter After Fermi,''
  arXiv:0908.4317 [hep-ph].
  %%CITATION = ARXIV:0908.4317;%%




\bibitem{Pamela}
  O.~Adriani {\it et al.}  [PAMELA Collaboration],
%  ``An anomalous positron abundance in cosmic rays with energies 1.5.100 GeV,''
  Nature {\bf 458}, 607 (2009),
  %%CITATION = NATUA,458,607;%%
  O.~Adriani {\it et al.},
%  ``A new measurement of the antiproton-to-proton flux ratio up to 100 GeV in
%  the cosmic radiation,''
  Phys.\ Rev.\ Lett.\  {\bf 102}, 051101 (2009).
  %%CITATION = PRLTA,102,051101;%%  
  
%\cite{ATIC}
\bibitem{ATIC}
  J.~Chang {\it et al.},
%  ``An Excess Of Cosmic Ray Electrons At Energies Of 300.800 Gev,''
  Nature {\bf 456} (2008) 362.
  %%CITATION = NATUA,456,362;%%

\bibitem{Fermi-LAT}
  A.~A.~Abdo {\it et al}\/ [Fermi LAB Collaboration],
 % ``Measurement of the Cosmic Ray e+ plus e- spectrum from 20 GeV to 1 TeV with
%  the Fermi Large Area Telescope,''
  Phys. Rev. Lett.  {\bf102}, 181101 (2009).
 % arXiv:0905.0025 [astro-ph.HE].
  %%CITATION = ARXIV:0905.0025;%%  
  
%\cite{Lillie:2007ve}
\bibitem{Lillie:2007ve}
  B.~Lillie, J.~Shu and T.~M.~P.~Tait,
  %``Kaluza-Klein Gluons as a Diagnostic of Warped Models,''
  Phys.\ Rev.\  D {\bf 76}, 115016 (2007).
%  [arXiv:0706.3960 [hep-ph]].
  %%CITATION = PHRVA,D76,115016;%%
  
%\cite{Alwall:2007st}
\bibitem{Alwall:2007st}
  J.~Alwall {\it et al.},
  %``MadGraph/MadEvent v4: The New Web Generation,''
  JHEP {\bf 0709}, 028 (2007).
%  [arXiv:0706.2334 [hep-ph]].
  %%CITATION = JHEPA,0709,028;%%  
    
\end{thebibliography}

%%%%%%%%%%%%%%%%%%%%%%%%%%%%%%%%%%%%%%%%%%%
%% The following lines show an example how to produce a bibliography
%% without the help of the BibTeX program. This could be used instead
%% of the above.
%%%%%%%%%%%%%%%%%%%%%%%%%%%%%%%%%%%%%%%%%%%

%\endinput

\end{document}